\documentclass[12pt]{iopart}
\usepackage{graphicx}
\usepackage{iopams}

\begin{document}

\newcommand{\nTHREEd}{n_{\rm{3D}}}
\newcommand{\nTHREEdTOF}{\widetilde{n}_{\rm{3D}}}
\newcommand{\nTHREEdMAX}{n_{\rm{3D}}(\mathbf{\vec{0}})}
\newcommand{\nTWOd}{n_{\rm{2D}}}
\newcommand{\nTWOdTOF}{\widetilde{n}_{\rm{2D}}}
\newcommand{\nCORE}{\overline{n}_{\rm{2D}}}

\title[Experimental study of the transport of BECs in a 1D random potential]{Experimental study of
the transport of coherent interacting matter-waves in a 1D random
potential induced by laser speckle}

\author{D. Cl\'ement, A. F. Var\'{o}n, J. A. Retter,
L. Sanchez-Palencia, A. Aspect and P. Bouyer}
\address{Laboratoire Charles Fabry de l'Institut d'Optique,
Centre National de la Recherche Scientifique et Universit\'e Paris
Sud 11, Batiment 503, Centre scientifique F91403 ORSAY CEDEX,
France, www.atomoptic.fr}

\begin{abstract}
We present a detailed analysis of the 1D expansion of a coherent
interacting matterwave (a Bose-Einstein condensate) in the
presence of disorder. A 1D random potential is created via laser
speckle patterns. It is carefully calibrated and the
self-averaging properties of our experimental system are
discussed. We observe the suppression of the transport of the BEC
in the random potential. We discuss the scenario of
disorder-induced trapping taking into account the radial extension
in our experimental 3D BEC and we compare our experimental results
with the theoretical predictions.
\end{abstract}

 \submitto{\NJP}

%\tableofcontents

\section{Introduction}

Disorder in quantum systems has been the subject of intense
theoretical and experimental activity during the past decades.
Since no real system is defectless, disordered systems are
actually more general than ordered ({\it e.g.} periodic) ones. In
solid state physics, disorder can result from impurities in
crystal structures, in the case of superfluid helium from the
influence of a porous substrate \cite{Glyde_PRL2000}, in the case
of micro-wave from alumina dielectric spheres randomly displaced
\cite{Dalichaouch_Nature1991,Chabanov_Nature2000}, in the case of
light from the transmission through a powder
\cite{Wiersma_Nature1997} and in ultracold atomic systems from the
roughness of a magnetic trap \cite{Esteve_PRA2004}. It is now well
established that even a small amount of disorder may have dramatic
effects, especially in 1D quantum systems
\cite{localizationbooks,vantiggelen1999}. The most famous and
spectacular phenomenon is certainly the localization and the
absence of diffusion of non-interacting quantum particles
\cite{Anderson_PR1958}, predicted in the seminal work of
P.W.~Anderson in the context of electronic transport. The quantum
phase diagrams of spin glasses \cite{spinglass} and
disorder-induced frustrated systems are other rich manifestations
of disorder.

In interacting systems, the situation is even richer and more
complex as a result of non-trivial interplays between kinetic
energy, interactions and disorder. This problem has attracted much
attention \cite{Fisher1989,disorderinteractions} but is still not
fully understood. In lattice Bose systems for example, it leads to
the formation of a Mott insulator and a Bose glass phase at zero
temperature \cite{Fisher1989}. A study of coherent transport of
two interacting particles also predicts a localization length
larger than the single-particle Anderson localization length
\cite{disorderinteractions}.

Recent progress in ultracold atomic systems has triggered a
renewed interest in quantum disordered systems where several
effects such as localization
\cite{Damski_PRL2003,Castin_PRL2005,martino2005,Sanchez-Palencia_PRA2005}
the Bose-glass phase transition
\cite{Damski_PRL2003,Roth_JOB2003,Fallani_arxiv2006} or the
formation of Fermi-glass, quantum percolating and spin glass
phases \cite{Sanpera_PRL2004,Ahufinger_2005} have been predicted
(for a recent review see \cite{Ahufinger_2005}). Ultracold atoms
in optical and magnetic potentials provide an isolated, defectless
and highly controllable system and thus offer an exciting (new)
laboratory in which quantum many-body phenomena at the border
between atomic physics and condensed matter physics can be
addressed \cite{natureinsights}. Controllable random potentials
can be introduced in these systems using several techniques. These
include the use of impurity atoms located at random positions of a
lattice \cite{impurity}, quasi-periodic potentials
\cite{Damski_PRL2003,Sanchez-Palencia_PRA2005,horak1998,guidoni},
optical speckle patterns
\cite{Lye_PRL2005,Clement_PRL2005,Fort_PRL2005} or random phase
masks \cite{Schulte_PRL2005}.

In this work, we experimentally investigate the transport
properties of an interacting Bose-Einstein condensate (BEC) in a
1D random potential. We use laser speckle to create a 1D repulsive
random potential along the longitudinal axis of cigar-shaped BEC.
To study the transport properties of the condensate in the random
potential, we observe the 1D expansion of the interacting
matter-wave in a magnetic waveguide, oriented along the axis of
the BEC. We demonstrate the suppression of transport
\cite{Clement_PRL2005} induced by the random potential and
carefully analyze the disorder-induced localization phenomenon
\cite{Laurent_paper}. In the regime that we consider (Thomas-Fermi
regime), the interactions play a crucial role for the observed
localization which turns out to be completely different from
Anderson localization. Compared to the other above-mentioned means
of creating disorder in ultracold atomic systems, this turns out
to have significant advantages. First, speckle patterns form
disordered potentials which are truly random with no long-range
correlation; second, they do not require two-species systems; and
third, their parameters (intensity and correlation functions) can
be shaped almost at will in 1D, 2D or 3D. Careful attention is
paid to the characterization of speckle patterns in connection to
ultracold atoms in the present work.

The article is organized as follow. We present the characteristics
of our random potential: its statistical properties and their
connection to experimental parameters in
\Sref{Section_speckle_field}, as well as methods to calibrate this
potential correctly in \Sref{Section_exp_implementation}. We
present the observation of inhibition of the expansion of an
interacting matter-wave in the random potential in
\Sref{Section_expansion}. We then discuss the disorder-induced
scenario proposed in \cite{Laurent_paper,Clement_PRL2005} and
present a detailed experimental analysis of this theoretical
scenario in \Sref{Section_exp_check_scenario}.

\section{Laser speckle: a controllable random potential for cold
atoms}\label{Section_speckle_field}

Shining a speckle pattern onto the BEC creates a random potential
for the atoms as they are subjected to an optical dipole potential
$V(\vec{r})$. This dipole potential is proportional to the
intensity $I(\vec{r})$ of the laser light and inversely
proportional to the detuning $\delta$ from the atomic transition:
\begin{equation}
\label{def_pot_dip} V(\vec{r}) = \frac{2}{3} \frac{\hbar\Gamma^2
}{8 I_{\rm{sat}}}  \frac{I(\vec{r})}{\delta},
\end{equation}
with $I_{\rm{sat}}=16.56$\,W\,m$^{-2}$ the saturation intensity of
the $D_2$ line of Rb$^{87}$, $\Gamma/2\pi=6.06\,$MHz the
linewidth, and the factor $2/3$ the transition strength for
$\pi$-polarized light. In this section, we present the main
characteristics of the random potential induced by a laser
speckle.

\subsection{What is a speckle field ?}

\begin{figure}[ht!]
\begin{center}
\includegraphics[width=13cm]{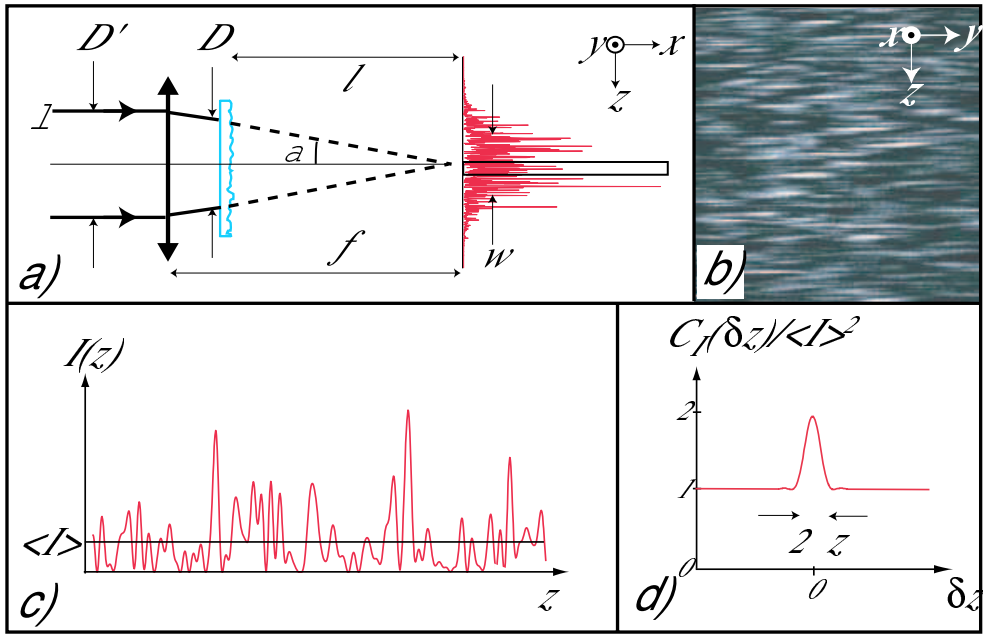}
\end{center} \caption{\textbf{a)} Experimental realization of the
speckle pattern. A laser beam of diameter $D'$ and wavelength
$\lambda$ is first focussed by a convex lens. The converging beam
of width $D$ is then scattered by a ground glass diffuser. The
transverse speckle pattern is observed at the focal plane of the
lens. The scattered beam diverges to an rms radius $w$ at the
focal plane. \textbf{b)} Image of an anisotropic speckle pattern
created using cylindrical optics to induce a 1D random potential
for the BEC along $Oz$. \textbf{c)} Zoom of the speckle pattern
(the boxed region of a). \textbf{d)} The intensity autocorrelation
function $C_I(\delta z)$ (defined in the text). Its width gives
the typical speckle grain size $\Delta z$.} \label{figure1}
\end{figure}

Laser speckle is the random intensity pattern produced when
coherent laser light is scattered from a rough surface resulting
in spatially modulated phase and amplitude of the electric field
(see \Fref{figure1}a) \cite{Goodman2005,Francon}. The
randomly-phased partial waves originating from different
scattering sites of the rough surface sum up at any spatial
position $\mathbf{r}$ leading to constructive or destructive
interferences. This produces a high-contrast pattern of randomly
distributed grains of light (see \Fref{figure1}b). A fully
developed speckle pattern is created when the rough surface
contains enough scatterers to diffuse all the incident light so
that there is no directly transmitted light. This requires the
phases acquired at each scatterer to be uncorrelated and uniformly
distributed between $0$ and $2\pi$. This is achieved by using a
rough surface whose profile has a variance which is large compared
with the wavelength of the light.

The real and imaginary parts of the electric field of the speckle
pattern are independent Gaussian random variables -- a consequence
of the central limit theorem \cite{Goodman_Stat_Optics}. Simple
statistics can be used to derive the properties of the resulting
intensity pattern which are related to that of the electric field:
(i) the first order one-point statistical properties which
correspond to the speckle intensity distribution, (ii) the
second-order two-point statistical properties which correspond to
the intensity correlation function and to the typical size of the
speckle grains. We show that all parameters of the speckle random
potential can be controlled accurately experimentally.

\subsection{Speckle Amplitude}\label{subsection_speckle_amplitude}

In a fully-developed speckle pattern, the sum of the scattered
partial waves results in random real and imaginary components of
the electric field whose distributions are independent and
Gaussian. Consequently the speckle intensity $I$ follows an
exponential law:
\begin{equation}\label{distributionI}
P(I)=\frac{1}{\left\langle I \right
\rangle}e^{-\frac{I}{\left\langle I \right \rangle}} .
\end{equation}
The amplitude of the speckle intensity modulation is defined by
its standard deviation $\sigma_I=\sqrt{\left\langle I^2 \right
\rangle-\left\langle I \right \rangle^2}$. From the intensity
distribution \eref{distributionI} it is easy to show that
$\sigma_I=\left\langle I \right \rangle$. The probability of a
speckle peak having an intensity equal to or greater than five
times the average intensity is less than 1\%. This will provide a
reasonable estimate of the highest speckle peaks (see
\Fref{figure1}c).

The average speckle intensity $\langle I \rangle$ is directly
related to the intensity of the incident laser beam and to the
diffusion angle of light scattered by the diffuser. This angle
increases as the minimum size of the scatterers on the diffuser
decreases, causing the divergence of the scattered beam to
increase, thereby reducing the average intensity. Reducing the
distance $l$ of the diffuser from the focal plane (see
\Fref{figure1}a) changes the average intensity of the speckle as
the laser beam diverges over a shorter distance, but without
changing the second-order statistical properties, as we will see
in the following.

\subsection{Speckle grain size and intensity correlation
function}\label{subsection_speckle_grain_size}

Roughly speaking a speckle pattern is a spatial distribution of
grains of light intensity with random magnitudes, sizes and
positions (see \Fref{figure1}b). The speckle grain size is
characterized by the width of the intensity autocorrelation
function (\Fref{figure1}d):
\begin{equation} \label{Intensity_Correlation}
 C_{I}(\delta \mathbf{r} ) =
\left\langle I(\mathbf{r}) I(\mathbf{r} + \delta \mathbf{r})
\right \rangle
\end{equation}
where $I(\mathbf{r})$ is the intensity at point $\mathbf{r}$ and
the brackets imply statistical averaging. This function can be
derived from the electric field statistics at the diffuser by
Fresnel/Kirchoff theory of diffraction \cite{Goodman2005,Francon}.
In the focal plane of the lens, assuming paraxial approximation,
the speckle electric field amplitude is the Fourier transform of
the electric field transmitted by the diffuser. Thus the
transverse autocorrelation function depends only on the linear
phase terms of the electric field. However, in the longitudinal
direction, the quadratic terms of the phase must be taken into
account, leading to a different scaling in the longitudinal
direction \cite{Goodman2005}.

For the simple case where the diffuser is illuminated by a uniform
rectangular light beam of width $D_Y$ and length $D_Z$ (as in our
experiment), the intensity correlation functions in the transverse
and longitudinal directions are respectively \cite{Goodman2005}:

\begin{eqnarray}\label{TransverseAuto}
  C_{I}^\bot(\delta y, \delta z) \;&=&\; \left < I \right > ^2
  \left [ 1 + f \left ( \frac{D_Y }{\lambda l} \delta y \right )
  f \left ( \frac{D_Z}{\lambda l} \delta z \right) \right ], \\
   C_{I}^\|(\delta x) \;&=&\; \left < I \right > ^2
  \left [ 1 + g \left ( \frac{D_Y^2}{\lambda l^2} \delta x \right )
  g \left ( \frac{D_Z^2}{\lambda l^2} \delta x \right) \right ].
  \label{LongitudinalAuto}
\end{eqnarray}
Here $f(u)=[\sin(\pi u)/ \pi u]^2$ is the Fourier transform of the
aperture, $g(u)=\frac{2}{u} \left [ C^2 \left ( \sqrt \frac{u}{2}
\right ) + S^2 \left ( \sqrt \frac{u}{2} \right ) \right ]$ where
$C(s)$ and $S(s)$ are the Fresnel cosine and sine integrals
respectively. \Eref{TransverseAuto} is valid in the far-field
regime, {\it i.e.} for $(\delta x^2 +\delta y^2) / l^2 << 1$. We
define the typical size of the speckle grains as the distance to
the first zero of the functions $\frac{C_I(\delta
\bold{r})}{C_I(0)}-1$ in each coordinate direction. We find the
following grain sizes for each of the three directions:
\begin{equation}
\Delta y=\lambda \frac{l}{D_Y} \ , \ \Delta z=\lambda
\frac{l}{D_Z} \ , \ \Delta x \simeq 7.6 \lambda \frac{l^2}{D_Y
D_Z} = \frac{7.6 \Delta y \Delta z }{\lambda}.
\end{equation}

An important point here is that aberrations of the optical setup
have no effect on the properties of the speckle observed in the
image plane \cite{Goodman2005}. It is interesting to note the
transverse speckle grain size corresponds to the diffraction
limit, {\it i.e.} it is controlled by the half-angle $\alpha=
D'/2f$ subtended by the illuminated area of the diffuser at the
observation point (see \Fref{figure1}a). As a consequence,
changing the distance of the diffuser relative to the lens does
not change the speckle grain size along $Oz$ since the angle
$\alpha$ remains constant. We also point out that the speckle
grain size in the focal plane of the lens is independent of the
size of the scatterers on the diffuser. We note that the
longitudinal grain size $\Delta x$ is related to the transverse
area $\Delta y \Delta z$ as the Rayleigh length of a gaussian beam
is related to the beam waist area (within a numerical factor).
Finally, we point out that when the half-angles in the transverse
planes $xOz$ and $xOy$ which determine the value $\Delta z$,
$\Delta y$ respectively are different, as in the experiment [see
\Fref{figure1}b], an anisotropic speckle pattern is created. As we
will explain in \Sref{Section_exp_implementation}, using an
anisotropic speckle pattern allows us to work with a 1D random
potential for the atoms.

\subsection{Self-averaging properties of a speckle pattern}
\label{subsection_self_averaging}

A 1D random potential $v(z)$, such as a speckle pattern, can be
defined by its statistical moments, ensemble averaged over the
disorder. Within this statistical definition, one can generate
many different realizations of the random potential.
Experimentally, this can be achieved by using different ground
glass diffusers or by shining different (uncorrelated) regions of
the speckle pattern onto the atom cloud. In principle,
experimental observations of cold atoms in a random potential will
depend on the microscopic details of each particular realization
of the random potential. Therefore, macroscopic transport
properties, which depend only on the statistics of the random
potential, should be extracted by \emph{ensemble averaging} over
many different realizations.

It is well known however that a spatially homogeneous (i.e.
infinite) disorder, without infinite range correlations, ensures
that all extensive physical quantities are ``self-averaged''
\cite{Lifshits_disorder_theory}. If a random potential is
self-averaging, we can obtain the statistical moments $m_i$ by
integrating a single realization of the random potential over an
infinite range: $m_i= \mathrm{lim}_{[D \rightarrow \infty]}
\frac{1}{D} \int_D \ dz \ v^{i}(z)$ for $i=1...\infty$. There is
then no need to average over many different realizations as each
gives exactly the same result. The fact that the spatial average
coincides with the statistical average is equivalent to the
well-known ergodic hypothesis in statistical mechanics, which
assumes that temporal mean is equal to the statistical average. In
experiments, studies are obviously carried out in \emph{finite}
systems, in which the self-averaging property is no longer
strictly valid. However, if the length $d$ of a 1D system is
sufficiently large (typically $d>>\Delta z$), the system will be
approximately self-averaging. More precisely, this approximation
will be valid if the statistical moments, evaluated over a finite
length $d$: $m_i(d) ~ = ~ \frac{1}{d} \int_{-d/2}^{d/2} dz \
v^{i}(z)$, yield values sufficiently close to the ensemble
averaged $m_i$.

In practice, it is useful to quantify the precision of this
approximation. This will identify under which circumstances it is
necessary to average experimental results over several
realizations of disorder, and under which circumstances it is
possible to assume self-averaging. For infinite systems
($d=\infty$), the self-averaging property implies
$\sigma^2_{m_i}(\infty)=<m_i(\infty)^2>-<m_i(\infty)>^2=0$, so the
calculation of the standard deviation $\sigma_{m_i}(d)$ of the
moment $m_i(d)$ gives a non-zero value which can be used to test
the extent to which a finite system is self-averaging. Rather than
calculating all the standard deviations, we will focus on just the
first and second-order deviations $\sigma_{m_1}(d)$ and
$\sigma_{m_2}(d)$. We will see later that the second-order moment
is a key parameter in our understanding of the transport
properties of the BEC.
\newline

We consider a 1D speckle potential $I(z)=\sigma_I v(z)$ with a
finite spatial correlation length $\Delta z$, where $v(z)$ is a
normalized speckle field: $< v(z)>=1$ and $<v^2(z)>=2$. For
simplicity let us approximate the auto-correlation function of the
speckle pattern to unity plus a Gaussian [this happens to be a
good approximation when the true auto-correlation function is
$C_v(z)=1+\frac{\sin[\pi z/ \Delta z]}{\pi z / \Delta z}$; see
\Fref{figure6}]. It is then possible to obtain a simple analytical
formula for $\sigma_{m_2}(d)$ [see
\ref{Appendix_calculation_sigma} and
\Eref{Eq:calcul_m1_carre}-\eref{Eq:relation_m1_m2}]. In the
asymptotic limit $d \gg \Delta z$, $\sigma_{m_2}(d)$ reduces to
[see \Eref{Eq:asympto_sigma_m2}-\eref{sigma_m2_zero}]:
\begin{equation}
\sigma_{m_2}(d)/\sigma_{m_2}(0) \ \simeq \ 0.959 \
\sqrt{\frac{\Delta z}{d}} \label{Eq:asympto},
\end{equation}
where $\sigma^2_{m_2}(0)=<v(z)^{4}>-<v(z)^2>^2=20$. As expected,
when the length $d$ of the medium tends to infinity, the system
becomes self-averaging and $\sigma_{m_2}(d)$ tends to zero. The
asymptotical convergence towards a self-averaging disorder is
however slow and scales as $\sqrt{\Delta z/d}$ where $d/\Delta z$
is the typical number of peaks present within the length $d$ of
the system. This scaling can be interpreted using discrete
variables. If we consider the amplitude $v(z_{\rm{k}})$ of the
random potential at the points $z_{\rm{k}}=\rm{k} \Delta z$, we
obtain a set of independent variables $\left ( v(z_{\rm{k}})
\right )_{\rm{k}=1..N}$ with the statistics of the speckle
intensity. Then the normalized spatial average $m_2(d)$ is a
normalized mean value over ${\rm N}=d/\Delta z$ independent
variables, which scales like $1/\sqrt{{\rm N}}$.

\begin{figure}[ht!]
\begin{center}
\includegraphics[width=7cm]{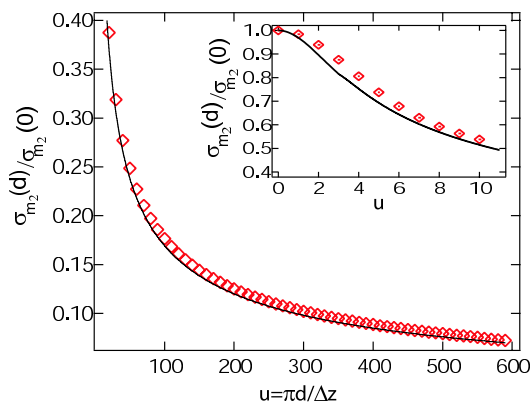}
\end{center}
\caption{\textit{Normalized standard deviation
$\sigma_{m_2}(d)/\sigma_{m_2}(0)$ as a function of the length $d$
of the system. Lozenges $\lozenge$ correspond to the numerical
calculation of $m_2(d)$ with the true auto-correlation function
$C_v(z)=1+\frac{\sin[\pi z/ \Delta z]}{\pi z / \Delta z}$. The
solid black line represents the analytical solution of
\Eref{Eq:calcul_m1_carre}-\eref{Eq:relation_m1_m2}. Inset: shows
plot in detail for small values of $d$. }\label{figure6}}
\end{figure}

In \Fref{figure6} we plot $\sigma_{m_2}(d)$ for the numerical
calculation using the true auto-correlation function
$C_v(z)=1+\frac{\sin[\pi z/ \Delta z]}{\pi z / \Delta z}$
(lozenges $\lozenge$) and the analytical solution with the
Gaussian auto-correlation function (solid black line). Both give
similar values for $\sigma_{m_2}(d)$. The asymptotic approximation
\Eref{Eq:asympto} is very good even for relatively small values of
$d/ \Delta z$: the deviation of \Eref{Eq:asympto} from the exact
solution of $\sigma_{m_2}(d)$ with the Gaussian approximation is
less than $1\%$ when the system is larger than six times the size
of the speckle grain. We note that the deviation from a
self-averaging system displayed by the first-order moment $m_1(d)$
is very similar to that of the second-order moment (see
\ref{Appendix_calculation_sigma}). For a typical number of peaks
$d/\Delta z$ larger than 100 as in our experiment, the difference
between the second order moment of a finite speckle pattern and
that of an infinite, self-averaging one is less than 10$\%$.

\section{Experimental implementation and characterization of the speckle
pattern}\label{Section_exp_implementation}
\subsection{Shining a speckle pattern onto the atomic cloud}
\label{subsection_shining_speckle}

\begin{figure}[ht!]
\begin{center}
\includegraphics[width=7.5cm,height=7.5cm]{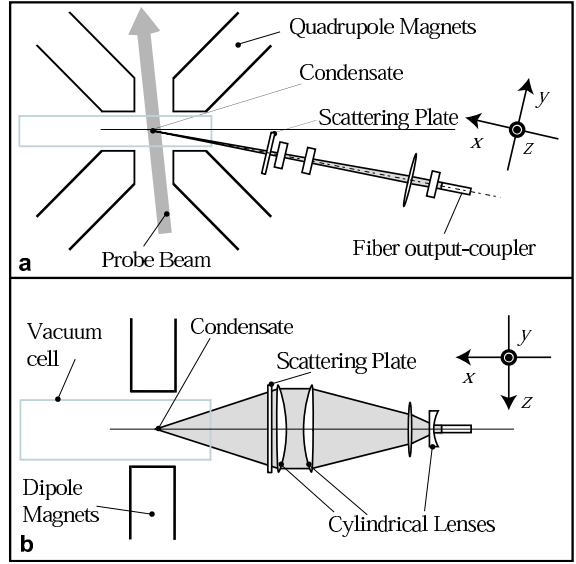}
\end{center}
\caption{\label{figure2}Optical setup used to create the speckle
potential. The BEC is at the focus of the lens system with its
long axis oriented along the $z$ direction. The two figures are
shown in the same scale. The beam incident on the diffuser has
different widths in the  $y$ and $z$ directions, which leads to
anisotropic speckle grains (see text). \textbf{a)} Side view.
\textbf{b)} Upper view.}
\end{figure}

In our experiment the random potential is superimposed on the
atoms by shining a laser beam through a ground glass diffuser as
shown in \Fref{figure2}. The atoms are located in the focal plane
of the lens (the observation plane in \Fref{figure1}), at a
distance $l=6\,$cm from the diffuser. The laser beam is derived
from a tapered amplifier, injected by a free-running diode laser
at $\lambda \sim 780$\,nm and fibre-coupled to the experiment. The
out-coupled beam is focused onto the condensate, the fibre
out-coupler and lenses being mounted on a single small optical
bench, aligned perpendicular to the long axis of the cigar-shaped
BEC.

The optical dipole potential $V(z)$ resulting from the speckle
pattern is (see \Eref{def_pot_dip}):
\begin{equation}\label{equ:V}
\langle V(z)\rangle=\sigma_V = \frac{2}{3} \frac{\hbar\Gamma^2}{8
I_{\rm{sat}}}  \frac{\sigma_I}{\delta}.
\end{equation}
In these experiments as in that of \cite{Clement_PRL2005}, we use
a blue-detuned light, ($\delta \gtrsim 0.15\,$nm), so the
potential is repulsive and the speckle grains thus act as barriers
for the atoms. This is in contrast to the case of a red-detuned
light ($\delta <0$) where the speckle grains act as potential
wells and where atoms could be trapped by the gaussian intensity
envelope of the laser beam. For the laser intensities used in
these experiments, the mean speckle potential $\sigma_V$ is always
smaller than the chemical potential of the initially trapped
condensate. We define the normalized amplitude of the random
potential $\gamma=\sigma_V / \mu_{\rm{TF}}$ relative to the
Thomas-Fermi chemical potential $\mu_{\rm{TF}}$ of the initially
trapped condensate. In our experiments, $\gamma$ is always smaller
than unity.

As explained in \Sref{subsection_speckle_grain_size}, we can
create an anisotropic speckle pattern by controlling the shape of
the laser beam incident on the diffuser. We use a set of
cylindrical optics such that in the $xOy$ plane (\Fref{figure2}a)
the out-coupled beam from the fibre is directly focussed onto the
atoms.  Thus along $Oy$ the height of the beam incident on the
diffuser is small, $D_Y= 0.95\,$mm, giving a speckle grain size
$\Delta y=49\,\mu$m. In the $xOz$ plane (\Fref{figure2}b), the
beam is first expanded before being focussed onto the atoms and
the horizontal size of the beam on the diffuser is $D_Z=55\,$mm,
giving a horizontal grain size $\Delta z=0.85\,\mu$m.  The
longitudinal grain size is $\Delta x = 406 \mu$m. With our
cigar-shaped Bose-Einstein condensates elongated along $Oz$, of
transverse radius $R_{\rm{TF}}=1.5\,\mu$m and longitudinal
half-length $L_{\rm{TF}}=150\,\mu$m along $Oz$, we have:
\begin{equation}
L_{\rm{TF}}\gg \Delta z  \ \mathrm{and} \ R_{\rm{TF}}\ll \Delta y
,\Delta x,
\end{equation}
and the speckle pattern can be considered as a
\emph{one-dimensional potential} for the condensate.

The scattered laser beam has a total power of up to 150\,mW and
diverges to rms radii $w_y$ and $w_z$ which are two orders of
magnitude larger than $R_{\rm{TF}}$ and $L_{\rm{TF}}$ respectively
at the condensate. Therefore the average intensity (the Gaussian
envelope) of the beam can be assumed constant over
the region where the atoms are trapped. %The calibration of the

\subsection{Calibration of the speckle grain size}

In principle, the size of the speckle grains $\Delta z$ can be
calculated from the parameters $l$ and $D$ of the optical system.
However, a large aperture cylindrical lens system is not stigmatic
and we have therefore calibrated the speckle grain size using
images from a CCD camera. The optical set-up is removed from the
BEC apparatus and the intensity distribution observed on a CCD
camera at the same distance $l$ as the atoms. Taking images with
various beam apertures $D_Z$, we determine the autocorrelation
function of the speckle patterns to obtain the grain size $\Delta
z$ that we plot versus $1/D_Z$. For speckle grain sizes larger
than the CCD camera pixels ($2\,\mu$m), we can fit the data with a
straight line, obtaining $\Delta z = 1.11(9) \times \lambda l/D_Z$
to be compared with the calculated grain $\Delta z = \lambda
l/D_Z$ with the paraxial assumption. The camera cannot resolve
speckle grains smaller than the pixel size and so we extrapolate
the fit to give the grain size corresponding to the aperture we
use: for $D_Z=55\,$mm, we obtain $\Delta z = 0.95(7)\,\mu$m. The
width of the auto-correlation function in the perpendicular axis
gives the experimental value $\Delta y = 54(1)\,\mu$m, leading to
the longitudinal size $\Delta x = 499 (38)\,\mu$m.

\subsection{Calibration of the speckle average intensity via light
shift measurements}\label{subsection_amplitude_calibration}

Obtaining a reliable value for a dipolar potential from a
photometric measurement of light intensity is notoriously
difficult. In our case, an additional problem arises due to the
strong focussing, entailing a strong variation of the intensity
along the beam axis $Ox$. The ideal method to calibrate the
dipolar potential is to use the atoms themselves as a sensor. This
potential is nothing else than the light-shift of the lower level,
$F=1$ in our case. In order to relate this light-shift to the
directly measured power of the laser that creates the speckle
random potential, we have used a measurement of the differential
light-shift of the $F=1 \rightarrow F=2$ hyperfine transition.

\begin{figure}[h]\begin{center}
\includegraphics[width=8cm]{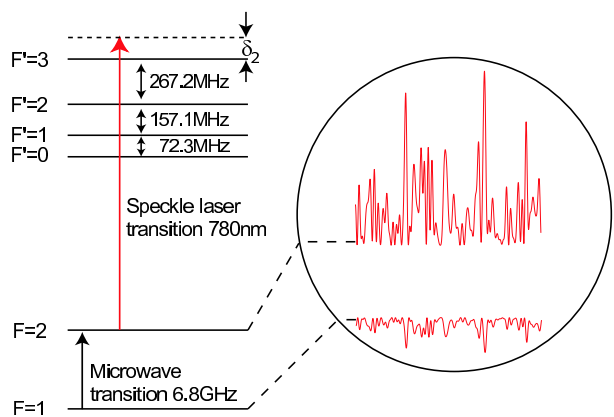}
\caption{\label{figure3}The speckle light at 780\,nm induces a
light shift in both $F=1$ and $F=2$.  Tuning the speckle laser
close to the $F=2$ level as shown creates a spatially varying
differential light-shift on the 6.8\,GHz transition.}
\end{center}\end{figure}

The condensate is magnetically trapped in the $|F\!=\!1,\,
m_F\!=\!-1\,\rangle$ sublevel (\Fref{figure3}). We use a microwave
frequency generator and antenna to drive the 6.8\,GHz $\sigma^{+}$
transition to the $|F\!=\!2,\,m_F\!=\!0\rangle$ sublevel. Atoms
coupled into this state are then lost from the trap. By monitoring
the number of atoms remaining in $|F\!=\!1,\, m_F\!=\!-1\,\rangle$
as a function of microwave frequency $f_{\rm{mw}}$, we obtain a
spectrum of this transition.

When the speckle laser is shone onto the atoms, both ground-states
$F=1$ and $F=2$ are light-shifted and the microwave spectrum is
modified. To produce a substantial differential light shift on the
transition, we must tune the speckle laser close to resonance for
either the $F=1$ or $F=2$ state. We chose to tune the laser close
to the $F=2\rightarrow F'=3$ transition frequency, to minimize
spontaneous scattering by the atoms trapped in the $F = 1$ level.
Then the speckle laser is sufficiently detuned from
$F=1\rightarrow F'$ transitions ($\delta_1 \sim 6.8$ GHz) that the
hyperfine structure of the upper $F'$ levels is not resolved for
this transition and the light-shift of the $|F\!=\!1,\,
m_F\!=\!-1\,\rangle$ sublevel is calculated using \Eref{equ:V}
[the transition strength is $2/3$ for $\pi$-polarized light]. For
the $F=2\rightarrow F'$ transitions, one has to take into account
the hyperfine structure of the excited state $F'$. The
$\pi$-polarized speckle laser beam couples the $|F\!=\!2,\,
m_F\!=\!0\,\rangle$ sublevel to $|F'\!=\!1,\, m_F\!=\!0\,\rangle$
and $|F'\!=\!3,\, m_F\!=\!0\,\rangle$ with the transition
strengths 1/15 and 3/5 respectively. For detunings close to
resonance with $F=2 \rightarrow F'=3$, the contribution to the
light-shift of the transition to the $F'=1$ sublevel is negligible
at the 1\% level. In this approximation we obtain the differential
light-shift of the transition:
\begin{eqnarray}\label{equ:lightshift}
\Delta E &=& \frac{\hbar \Gamma^2}{8 I_{\rm{sat}}} \ I
\left (\frac{3}{5} \frac{1}{\delta_2}- \frac{2}{3} \frac{1}{\delta_1}\right )  \\
\nonumber &\simeq& \frac{3}{5}\frac{\hbar \Gamma^2}{8
I_{\rm{sat}}} \frac{I}{\delta_2} \quad \rm{for}\ \delta_2 \ll
\delta_1 \nonumber
\end{eqnarray}
where $\delta_1$ is the detuning relative to the $F=1\rightarrow
F'=3$ transition, $\delta_2$ is the detuning relative to the
$F=2\rightarrow F'=3$ transition ($\delta_1=\delta_2-6.835$\,GHz).
Note that the laser is always red-detuned from $F=1$ by about $
-6.8\,$GHz, so creates an attractive speckle potential for the
atoms, while it is red or blue-detuned for the $F=2$ state. Atoms
transferred to $F=2$ by the microwave pulse will be rapidly lost
due to near resonant spontaneous scattering.

\begin{figure}[h]\begin{center}
\includegraphics[width=16cm]{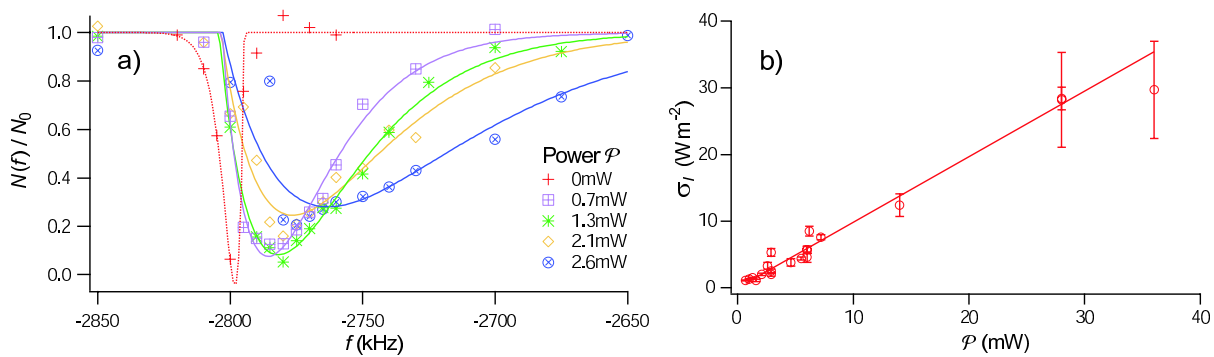}
\caption{\label{figure4} \textbf{a)} Fraction of atoms remaining
in $F=1$ after a 5\,ms pulse of microwaves at frequency
$f_{\rm{mw}}=6.8\rm{GHz}+f$, for speckle laser powers
$\mathcal{P}$ indicated and fixed (blue) detuning
$\delta=15\,$MHz. Increasing the speckle laser power broadens the
spectra to lower microwave frequencies. (Frequency $f$ is
indicated relative to the unshifted transition frequency at
6.834683\,GHz.) The solid lines represent fits to \Eref{equ:fit}.
\textbf{b)} $\sigma_I$ versus speckle laser power $\mathcal{P}$.
The fit gives $\sigma_I/\mathcal{P}=1.0(1) \times 10^3
\rm{m}^{-2}$.}
\end{center}\end{figure}

Spectra obtained for a condensate in the absence of the speckle
potential (red crosses $+$ on \Fref{figure4}a) have a width of
$\simeq 15\,$kHz and are shifted by $f_B=f_{\rm{mw}}-6.8$GHz
$\simeq -2800\,$kHz from the $F=1\rightarrow F=2$ transition
frequency. This shift and this width are due to the Zeeman effect
on the magnetic $|F\!=\!1,\, m_F\!=\!-1\,\rangle$ sublevel: the
minimum magnetic field $B_0$ of the Ioffe trap shifts the
frequency transition by $f_B= g_F \mu_B B_0 /h$ and the curvature
of the magnetic trap over the region of the condensate broadens
the spectrum towards lower frequencies. When the speckle laser is
shone on the atoms, different atoms experience different
light-shifts due to the spatial modulations of intensity in the
speckle pattern. The spectrum is therefore inhomogeneously
broadened due to the range of light intensities, as shown in
\Fref{figure4}a. For these measurement we used speckle intensities
of $\langle I \rangle \lesssim 0.3\,$mW.\,cm$^{-2}$ and detunings
$\delta_2$ from -15\,MHz to -500\,MHz.

\begin{figure}[h]\begin{center}
\includegraphics[width=7cm]{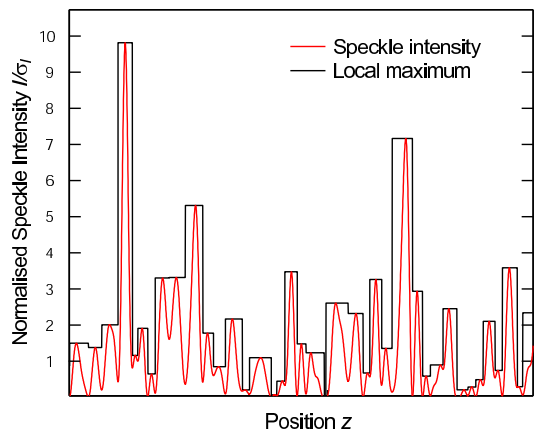}
\caption{\label{figure5} Plot of the `nearest local maxima'
effective potential (black line) of one realization of the speckle
pattern (red line). We use this simulation to obtain the
probability distribution of the `nearest local maxima' $P'(I)$.}
\end{center}\end{figure}

In order to calibrate the dipolar potential and to extract the
average intensity $\sigma_I$ from these experimental spectra, we
have developed a simple model. Since the broadening of the spectra
due to the variation of the speckle intensity is very large
compared with the Zeeman broadening we neglect the latter.
Approximating to a constant density profile, we use the statistics
of the speckle intensity distribution to model the evaporation.
Since the mean potential $\sigma_V$ is typically 100 times greater
than the chemical potential of the condensate, we assume that the
atoms are located essentially at the maxima of the speckle
intensity peaks (minima of the trapping potential). The number of
atoms remaining after application of the microwave pulse at
frequency $f_{\rm{mw}}=6.8\,$GHz$+f$ is then:
\begin{equation}\label{equ:Nf}
N(f)= N_0 \left [ 1- \alpha \frac{ \Delta f}{3 \Gamma^2 /80 \pi
I_{\rm{sat}} \delta_2} P'(I(f))\right ]= N_0 \left [ 1-
\mathcal{A} P'(I(f))\right ]
\end{equation}
where $\alpha$ is the coupling efficiency of the microwave knife
and $\Delta f$ the frequency width coupled by the microwave knife,
$I(f)$ is the intensity resonant with the frequency $(f-f_B)$,
{\it i.e.}
\begin{equation}\label{Eq:f_I_relation}
h (f-f_B)=\frac{3 \hbar \Gamma^2 I(f)}{40 I_{\rm{sat}} \delta_2}.
\end{equation}
In \Eref{equ:Nf} $P'(I)$ is the distribution of `nearest local
maxima' of intensity given by:
\begin{equation}\label{equ:Pmax}
P'(I')= \frac{4 I' \exp(-2 I'/ \bar{I'})}{ \bar{I}^{'2}}
\end{equation}
where $\bar{I}^{'} = 1.89 \sigma_I$ is the average value of the
distribution of intensity maxima. \Eref{equ:Pmax} is obtained by
simulations of the speckle distribution in which the intensity $I$
at each point of the speckle random potential was replaced by the
intensity at the nearest maximum $I'$ (see \Fref{figure5}).

We finally obtain from \Eref{equ:Nf}-\eref{equ:Pmax}:
\begin{equation}\label{Eq:Nf_bis}
N(f)=N_0 \left [ 1- \mathcal{B} \frac{(f-f_B)}{(1.89 \sigma_I)^2}
\exp \left ( -2 \frac{ f-f_B}{1.89 (f_{\sigma_I}-f_B)} \right )
\right ]
\end{equation}
where $\mathcal{B}=320 \pi \mathcal{A} I_{\rm{sat}} \delta_2 / 3
\Gamma^2$.

In the experiment, we measure the power $\mathcal{P}$ at the input
of the optical setup that creates the speckle random potential.
The aim of the calibration is thus to relate $\mathcal{P}$ to the
average intensity $\sigma_I$ of the random potential on the atoms.
In order to extract $\sigma_I$ from the experimental data we fit
the spectra in \Fref{figure4}a with a function similar to
\Eref{Eq:Nf_bis}:
\begin{equation}\label{equ:fit}
N(f)=N_0 \left [ 1- \mathcal{C} (f-f_B) \exp \left ( -2 \frac{
f-f_B}{1.89 (f_{\sigma_I}-f_B)} \right ) \right ]
\end{equation}
where $\mathcal{C}$, $f_B$ and $f_{\sigma_I}$ are fitting
parameters. Plotting the fitted values of $f_{\sigma_I}$ versus
$\mathcal{P}/\delta_2$ for red- and blue-detuned light we then
obtain a more accurate value of $f_B$ than that obtained by
fitting \Eref{equ:fit} to each individual spectrum. Using
\Eref{Eq:f_I_relation} we obtain $\sigma_I$ from
$f_{\sigma_I}-f_B$ and plot $\sigma_I$ versus $\mathcal{P}$ in
\Fref{figure4}b. We obtain $\sigma_I/\mathcal{P}=1.0(1) \times
10^3 \rm{m}^{-2}$ as the calibration constant relating the power
$\mathcal{P}$ to the average speckle intensity at the atoms
$\sigma_I$.

\section{Expansion in a 1D waveguide : a study of transport
properties of a Bose-Einstein condensate in a speckle random
potential}\label{Section_expansion}

\subsection{Production of a condensate of $^{87}$Rb atoms\label{subsection_BEC}}

We produce a Bose-Einstein condensate of $^{87}$Rb atoms in the
$|F=1,m_F=-1>$ hyperfine state. The design of our iron-core
electromagnet allows us to create an elongated Ioffe-Pritchard
magnetic trap with axial and radial frequencies of $\omega_z =
2\pi \times 6.70(7)$ Hz and $\omega_{\perp} = 2 \pi \times 660(4)$
Hz respectively. The magnetic trap is loaded from a
magneto-optical trap (MOT) and the atom cloud is cooled down to
quantum degeneracy (BEC) using a radio-frequency (rf) evaporation
ramp. Typically, our BECs comprise 3.5 $\times 10^5$ atoms and are
characterized by a chemical potential $\mu_{\rm{TF}}/2 \pi \hbar
\sim 4.6$ kHz and Thomas-Fermi half length $L_{\rm{TF}} = 150$
$\mu$m and radius $R_{\rm{TF}}=1.5$ $\mu$m. Further details of our
experimental apparatus are presented in \cite{Boyer_PRA2000}.

The large aspect ratio of the trap is of primary importance for
the experiments described in this article. As stated in
\Sref{subsection_shining_speckle} the large anisotropy of the
speckle grains ($\Delta x=499(38)\,\mu$m, $\Delta y=54(1)\,\mu$m
and $\Delta z=0.95(7)\,\mu$m) help us to obtain a 1D random
potential for the atoms. Yet to work with true 1D random
potentials, {\it i.e.} $L_{\rm{TF}} / \Delta z \gg 1$ and
$R_{\rm{TF}} / \Delta x \ll 1$, an order of magnitude difference
between the sizes $L_{\rm{TF}}$ and $R_{\rm{TF}}$ of the BEC is
needed. With the aspect ratio of our magnetic trap we have
$L_{\rm{TF}} / \Delta z \simeq 158$ and $R_{\rm{TF}} / \Delta x
\simeq 0.03$.

\subsection{Axial expansion: opening the longitudinal trap}

To study the coherent transport of the BEC in the random
potential, we observe the longitudinal expansion of the condensate
in a long magnetic guide. The setup of our magnetic trap allows us
to control almost independently the longitudinal and transverse
trap frequencies by changing the currents in the axial and radial
excitation coils. By reducing the axial confinement without
modifying the transverse confinement, we create a 1D magnetic
waveguide for the condensate. Repulsive inter-atomic interactions
drive the longitudinal expansion of the BEC along this guide.

Reducing the current in the axial excitation coils reduces both
the longitudinal trap frequency and the minimum value of the
magnetic field. If the minimum magnetic field crosses zero, atoms
can undergo Majorana spin-flips from the trapped hyperfine state
$|F=1,m_F=-1>$ to non-trapped hyperfine states and are then lost
from the trap. Therefore we monitor the atom number as a function
of the axial current in order to determine the current at which
the magnetic field crosses zero. This zero-crossing defines a
lower limit for the axial current and so we reduce the axial trap
frequency $\omega_z/2 \pi$ to a final value slightly above this
limit. Since we cannot reduce the axial field curvature strictly
to zero, a small longitudinal trapping remains. By observing
dipole and quadrupole oscillations (at frequencies $\omega'_z$ and
$\sqrt{2/5} \ \omega'_z$ respectively) in the magnetic waveguide,
we measured $\omega'_z/2 \pi = 1.10(5)$ Hz for the residual
trapping frequency in the guide.

Opening the trap abruptly induces atom loss and heating of the
atom cloud, therefore the trap is ramped over 30\,ms to avoid
these processes. Once the current in the axial coils has reached
its final value we have a BEC of N $\sim 2.5 \times 10^5$ --– $3
\times 10^5$ atoms in the magnetic guide.
\newline

To perform the experiment in the presence of the random potential,
we use the following procedure. After creating the condensate of
$^{87}$Rb atoms we shine the random potential onto the atoms and
wait 200 ms for the BEC to reach equilibrium in the combined
initial trap and disorder potential. We then open the longitudinal
confinement, switch off the evaporation RF knife and the BEC
expands in the 1D waveguide in the presence of disorder due to
repulsive interactions. We turn off all remaining fields
(including the random potential) after a total axial expansion
time $\tau$ (which includes the 30 ms opening time of the axial
trap) and wait a further $t_{\rm{tof}}=15$ ms of free fall before
imaging the atoms by absorption.

\subsection{Image analysis and longitudinal density profiles}
\label{subsection_image_analysis}

In our experiment we obtain quantitative information about the
atom cloud by taking absorption images after a time-of-flight
$t_{\rm{tof}}$=15 ms. Absorption imaging effectively integrates
the atomic density along the direction of the imaging beam $Oy$,
such that we measure the 2D density after time-of-flight
$\nTWOd(x,z,t_{\rm{tof}})$.
\newline

In the harmonic trap ($\tau=0$) without disorder, the Thomas-Fermi
condition is fulfilled and this justifies the use of the scaling
theory \cite{Scaling_factors1996}. During the time-of-fligth
$t_{\rm{tof}}$, the atom cloud expands with the scaling factors
$\lambda_{\perp}(t_{\rm{tof}})$ and $\lambda_z(t_{\rm{tof}})$ in
the radial and longitudinal directions respectively. In our
elongated trap $\lambda_z(t_{\rm{tof}}) \simeq 1$ and, after a
time-of-flight $t_{\rm{tof}}$, we have:
\begin{equation}\label{Eq:n2D_tof}
\nTWOd(x,z,t=t_{\rm{tof}})=\frac{1}{\lambda_{\perp}(t_{\rm{tof}})}
\ \nTWOd \left [ \frac{x}{\lambda_{\perp}(t_{\rm{tof}})},z,t=0
\right ].
\end{equation}

In the magnetic waveguide, the radial frequency $\omega_{\perp}$
is unchanged. When the longitudinal expansion is stopped, {\it
i.e.} when there is no longitudinal kinetic energy, the energy of
atom cloud is also totally transferred to the transverse degree of
freedom during $t_{\rm{tof}}$ with the scaling factor
$\lambda_{\perp}$ of the initial trap. Then \Eref{Eq:n2D_tof} is
still valid to describe the expansion of the condensate from the
magnetic waveguide during the time-of-flight. Therefore the 2D
density before time-of-flight is
$\nTWOd(x,z,\tau)=\lambda_{\perp}(t_{\rm{tof}}) \
\nTWOd(\lambda_{\perp}(t_{\rm{tof}})x,z,\tau+t_{\rm{tof}})$ where
$\tau$ is the time spent by the BEC in the waveguide. In the
waveguide in presence of a random potential, the 3D density can be
written as:
\begin{equation}
\nTHREEd(x,y,z,\tau) = \frac{1}{g} \left [ \mu(\tau)-m
\omega_{\perp}^2 x^2 /2-m \omega_{\perp}^2 y^2 /2-V(z) \right ],
\end{equation}
where $\mu(\tau)$ is the chemical potential in presence of the 1D
random potential $V(z)$, $g=\frac{4 \pi \hbar^2 a}{m}$ is the
interaction parameter and $a$ is the scattering length. Then the
2D density is:
\begin{equation}\label{Eq:n2D_mu}\fl
\nTWOd(x,z,\tau)= \int \ dy \ \nTHREEd(x,y,z,\tau) = \frac{4
R_{\rm{TF}}}{3 g \sqrt{\mu_{\rm{TF}}}} \left [ \mu(\tau) -m
\omega_{\perp}^2 x^2 /2- V(z) \right ]^{3/2}.
\end{equation}
In particular, assuming the amplitude of the random potential
$<V>=\sigma_V$ is small compared to the chemical potential $\mu$,
$\sigma_V \ll \mu$, we have $\nTWOd(x=0,z,\tau) \simeq \frac{4
R_{\rm{TF}} \mu(\tau)^{3/2}}{3 g \sqrt{\mu_{\rm{TF}}}}$ up to
first order in $\sigma_V / \mu$.
\newline

From these 2D images, we extract the ``longitudinal density
profile'' $\nTWOd(x=0,z,\tau)$ and use it to calculate the rms
length $L$ and the centre-of-mass position of the expanding
condensate in the absence or presence of the random potential.

\subsection{Expansion of the BEC and time evolution of rms length $L$}
\label{subsection_expansion_and_L}

\begin{figure}[ht!]
\begin{center}
\includegraphics[width=16cm]{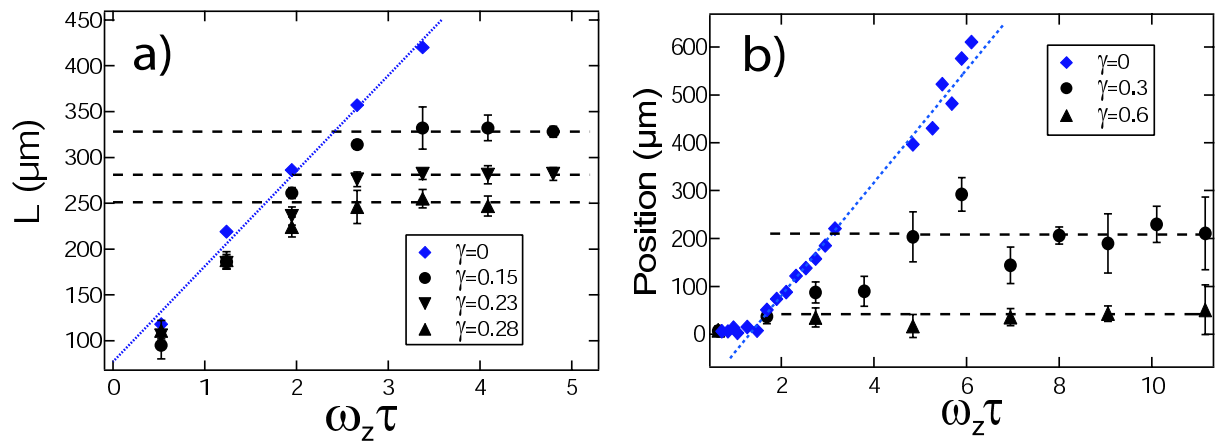}
\end{center}
\caption{\textit{\textbf{a)} Time evolution of the rms size L of
the expanding Bose-Einstein condensate in the 1D magnetic guide in
presence of a 1D random potential with amplitude $\gamma =
\sigma_V / \mu_{\rm{TF}}$. Error bars represents standard
deviation over 5 realizations of the speckle pattern (dashed lines
are guides to the eye) ; \textbf{b)} Time evolution of the centre
of mass position for the different values of
$\gamma$.}\label{figure7}}
\end{figure}

We measure the rms size $L$ and the centre-of-mass position of the
condensate as a function of the longitudinal expansion time $\tau$
and plot these quantities as a function of $\omega_z \tau$ in
\Fref{figure7}. The expansion of the condensate in the 1D
waveguide without disorder ($\gamma$=0) is linear as predicted by
the scaling theory \cite{Scaling_factors1996}. When the 1D random
potential is added ($\gamma$=0.15, 0.23, 0.28) the expansion is
reduced and eventually stops as reported in
\cite{Clement_PRL2005}. The dashed lines in \Fref{figure7}a
indicate $L^{\rm{f}}$, the final rms size of the condensate once
it stops expanding. The data in \Fref{figure7}a indicate that the
larger the normalized amplitude of the random potential compared
to the initial chemical potential ($\gamma$), the shorter the
final rms length $L^{\rm{f}}$ of the condensate.

\Fref{figure7}b shows the centre-of-mass position of the BEC as a
function of axial expansion time $\tau$. In the absence of the
random potential, the condensate acquires a centre-of-mass
velocity of 2.8(1) mm\,s$^{-1}$, due to a magnetic `kick' during
the longitudinal opening of the trap. We observe that applying a
small amplitude random potential also inhibits this centre-of-mass
motion. The displacement of the condensate decreases with
increasing amplitude $\gamma$ as shown in \Fref{figure7}b. Each
point in \Fref{figure7} is obtained by averaging the experimental
results over 5 different realizations of the speckle pattern. The
error bars represent the corresponding standard deviations. We
find that these standard deviations are not larger than the
shot-to-shot deviation observed using a single realization of the
random potential. We therefore claim that this system is
self-averaging within our experimental resolution. Further
justification is presented in \Sref{Subsection_average_density}.
This self-averaging property of our system allows us to measure
transport properties without averaging over many realizations of
the disorder, which is an important practical advantage.
\newline

\begin{figure}[ht!]
\begin{center}
\includegraphics[width=10cm]{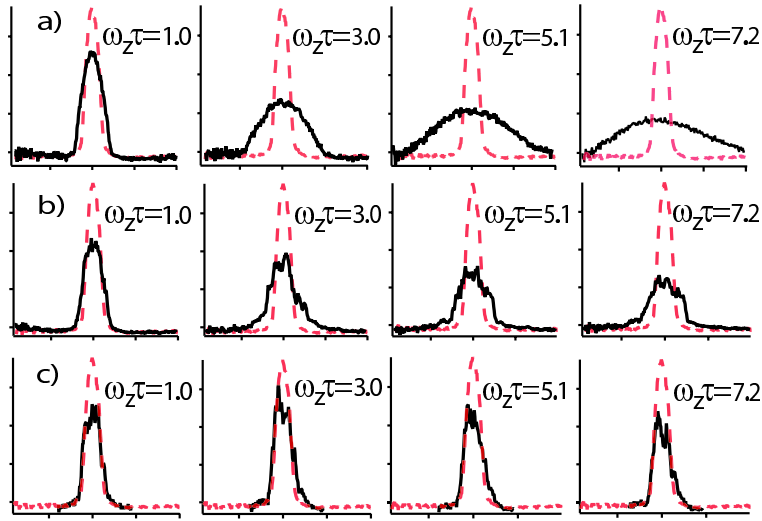}
\end{center}
\caption{\textit{Longitudinal density profiles of the expanding
BEC for times $\omega_z \tau=$1.0, 3.0, 5.1 and 7.2 with
\textbf{a)} $\gamma$=0, \textbf{b)} $\gamma$=0.15 and \textbf{c)}
$\gamma$=0.3. The dotted (red) line is the longitudinal profile in
the initial trap ($\tau=0$) in the absence of
disorder.}\label{figure8}}
\end{figure}

The suppression of transport of the expanding matter-wave also
appears clearly on the longitudinal density profiles obtained in
the experiments. We plot in \Fref{figure8} the time evolution of
the longitudinal density profiles for different values of the
amplitude $\gamma$ of the random potential. The dotted red profile
on every graph represents the longitudinal profile before
expansion ($\tau=0$) in the absence of the random potential. This
is the usual inverted parabola for a harmonically trapped
Bose-Einstein condensate in the Thomas-Fermi regime. During the
expansion without disorder (see \Fref{figure8}a), the shape
remains an inverted parabola with the rms size increasing as
expected from the scaling theory \cite{Scaling_factors1996}. When
the random potential is added (see \Fref{figure8}b and c), the
longitudinal density profile changes with time. Eventually it
reaches a stationary shape (corresponding to the stationary rms
size on \Fref{figure7}a) with two main characteristics: (i) a
constant central density (with random spatial modulations) and
(ii) steep edges demarcating this central region.
\newline

Our experimental results clearly show the suppression of transport
of a coherent matter-wave induced by a random potential
\cite{Clement_PRL2005}. Although phenomenologically similar to a
single particle Anderson localization (AL), we have argued in
Refs.\cite{Laurent_paper,Clement_PRL2005} that, in the mean-field
regime of our experiment where interactions are weak but
interaction energy dominates over the kinetic energy, the physics
strongly changes compared with that of AL with non-interacting
bosons. In the following we investigate experimentally the
scenario of disorder-induced trapping of an interacting
Bose-Einstein condensate in the mean field regime proposed in
Refs.\cite{Laurent_paper,Clement_PRL2005}.

\section{Experimental characterization of the disorder-induced trapping scenario}
\label{Section_exp_check_scenario}

\subsection{The disorder-induced trapping scenario of an elongated BEC}

The disorder-induced trapping scenario proposed in
\cite{Laurent_paper,Clement_PRL2005} describes the expansion of a
1D interacting matter-wave in a 1D random potential in a regime
where interactions dominate over the kinetic energy. The dynamics
of the BEC is governed by three kinds of energy: the amplitude of
the random potential, the kinetic energy of the atoms and the
inter-atomic interaction energy. The relative importance of each
of these three energy contributions depends on the local density
of the BEC. At the centre of the condensate where the density
remains large, interactions play a crucial role and the kinetic
energy is negligible. In contrast, the wings are populated by fast
atoms with a low density and therefore the kinetic energy
dominates over the interaction energy. Let us briefly describe
what happens in each of these regions.
\newline

In the wings, the kinetic energy dominates over the interaction
energy. The fast, weakly interacting atoms which populate the
wings undergo multiple reflections and transmissions on the
modulations of the random potential (see numerical simulations in
\cite{Laurent_paper,Clement_PRL2005}). Trapping finally results
from a classical reflection on a large modulation of the random
potential. The density in the wings of the disorder-trapped BEC is
too small to be measured by absorption imaging in our experiment.

We will now focus on the behaviour in the centre of the BEC.
Following the definitions of \cite{Laurent_paper}, we arbitrarily
delimit the core of the BEC as half the size of the initial
condensate: $-L_{\rm{TF}}/2 <z<L_{\rm{TF}}/2$. During the
expansion of the condensate in the waveguide, the density at the
centre of the cloud slowly decreases. It is thus possible to
define a quasi-static effective chemical potential
$\mu_{\rm{eff}}(\tau)$ for the core of the BEC after an expansion
time $\tau$ in the magnetic guide. The kinetic energy is small,
the time evolution is quasi-static and the healing length
$\xi=0.11~\mu$m remains much smaller than the correlation length
of the speckle potential $\Delta z=0.95~\mu$m: thus the
Thomas-Fermi approximation is valid. As a consequence, the random
potential modulates the density and the calculation of
$\mu_{\rm{eff}}(\tau)$ requires averaging over the length of the
core. The effective chemical potential $\mu_{\rm{eff}}(\tau)$ is
simply the sum of the interaction energy and the random potential
energy and is defined as
$\mu_{\rm{eff}}(\tau)=\frac{1}{L_{\rm{TF}}}
\int_{-L_{\rm{TF}}/2}^{L_{\rm{TF}}/2} dz \ [ g \
\nTHREEd(x=0,y=0,z,\tau) \ + \ V(z) ]$. The rapid lost of the
overall parabolic shape during the initial expansion of the BEC in
the random potential (see \Fref{figure8} and \cite{Laurent_paper})
justifies this expression for $\mu_{\rm{eff}}(\tau)$ with no
longitudinal magnetic trapping term. The effective chemical
potential $\mu_{\rm{eff}}(\tau)$ slowly decreases with the density
during the axial expansion time $\tau$ and eventually drops to a
value smaller than the amplitude of some peaks of the speckle
potential. Once this situation is reached, the condensate is
trapped in the region between these peaks and it fragments
\cite{Laurent_paper,Clement_PRL2005}. The criterion for trapping
the core of the BEC is the existence of two large modulations of
the random potential equal or greater than the effective chemical
potential $\mu_{\rm{eff}}$. Below, we adapt the calculations of
\cite{Laurent_paper} for 1D BECs  to take into account the
transverse extension of our 3D atom cloud. We note that the
calculations of \cite{Laurent_paper} assume a random potential
with $<V>=0$ so that the effective chemical potential reduces to
the interaction term: $\mu(\tau)=\mu_{\rm{eff}}(\tau) - <V>$
\footnote{This expression for $\mu(\tau)$ is strictly valid if
$<V>=\frac{1}{L_{\rm{TF}}} \int_{-L_{\rm{TF}}/2}^{L_{\rm{TF}}/2}
dz \ V(z)$, {\it i.e.} that our system is self-averaging on the
first-order moment $m_1$. Given our experimental setup, this
approximation is valid : $\sigma_{m_1}(L_{\rm{TF}}) \simeq 9 \%$
inferior to experimental uncertainties $\simeq 15\%$.}. However,
none of the physics of our experiment is lost by making this
adjustment with the time-independent energy $<V>$. In the
following we will use this new effective chemical potential
$\mu(\tau)$.
\newline

As the speckle potential is truly a 1D potential we can write the
condition of fragmentation in our 3D experimental BEC in the same
way as it is done in \cite{Laurent_paper} for a 1D BEC: the
picture developed for 1D BECs holds for the experimental 3D
condensates with the effective chemical potential $\mu(\tau)$
defined here. On the one hand if the BEC is fragmented at the
centre ($r^2=x^2+y^2=0$) then the condition for fragmentation
holds along the $x$-axis and $y$-axis since the density decreases
as $|r|$ increases. On the other hand when the condition of
fragmentation is not fulfilled at the centre of the BEC ($r=0$)
the BEC expands and there is no disorder-induced trapping.

The number of peaks in the core of the BEC with energy greater
than the effective chemical potential $\mu(\tau)$ is given by:
\begin{equation}
N_{\rm{peaks}}(V \geq \mu(\tau)) \simeq 0.94 \left (
\frac{L_{\rm{TF}}}{\Delta z} \right ) \exp \left [ -0.75
\frac{V}{\sigma_V} \right ]
\end{equation}
The condition of fragmentation is $N_{\rm{peaks}}=2$, and leads to
a relation between the final effective chemical potential
$\mu^{\rm{f}}$ once the core of the BEC is trapped and the
characteristic parameters $\sigma_V$ and $\Delta z$ of the speckle
potential. For small values of $\gamma=\sigma_V / \mu_{\rm{TF}}$
we obtain :
\begin{equation}
\mu^{\rm{f}} \simeq \frac{\mu_{\rm{TF}}}{ 0.75} \ \gamma \
\mathrm{ln} \left ( \frac{ 0.47 L_{\rm{TF}}}{ \Delta z} \right ),
\label{gamma_dependence}
\end{equation}
the logarithmic term reflecting the exponential probability
distribution of intensity of the speckle pattern (see
\Sref{subsection_speckle_amplitude}) and $\Delta z$ the
second-order statistics of our speckle potential.

Since the effective chemical potential $\mu(\tau)$ of the core of
the BEC decreases during the expansion, it cannot be larger than
the initial value $\mu(\tau=0)=\mu^{\rm{i}}=\frac{g}{L_{\rm{TF}}}
\int_{-L_{\rm{TF/2}}}^{L_{\rm{TF/2}}} \ dz \ \nTHREEd(0,0,z)$.
Integration over the core gives:
\begin{equation}
\mu^{\rm{i}} =\frac{11 \mu_{\rm{TF}}}{12} \simeq 0.92
\mu_{\rm{TF}} \label{saturated_mu}
\end{equation}
which thus provides an upper value for $\mu^{\rm{f}}$.
\newline

In order to compare the experiments with this scenario we have to
extract the effective chemical potential $\mu(\tau)$ in the core
of the condensate from the data (as detailed in
\Sref{subsection_image_analysis}). We extract the mean density
from our longitudinal profiles by averaging the density over the
core of the condensate $ \nCORE(\tau) = \frac{1}{L_{\rm{TF}}}
\int_{-L_{\rm{TF}}/2}^{L_{\rm{TF}}/2} dz \ \nTWOd(x=0,z,\tau)$.
Then the experimental effective chemical potential is [see
\Eref{Eq:n2D_mu})]:
\begin{equation}
\mu(\tau)= \mu_{\rm{TF}}^{1/3}\left ( \frac{3 g \nCORE(\tau)}{4
R_{\rm{TF}}}\label{mu_nCORE} \right ) ^{2/3},
\end{equation}
so that $\mu(\tau)$ can be directly extracted from the measured
density $\nTWOd$.

\subsection{Measurement of the average density in the core of the elongated
condensate}\label{Subsection_average_density}

\begin{figure}[ht!]
\begin{center}
\includegraphics[width=17cm]{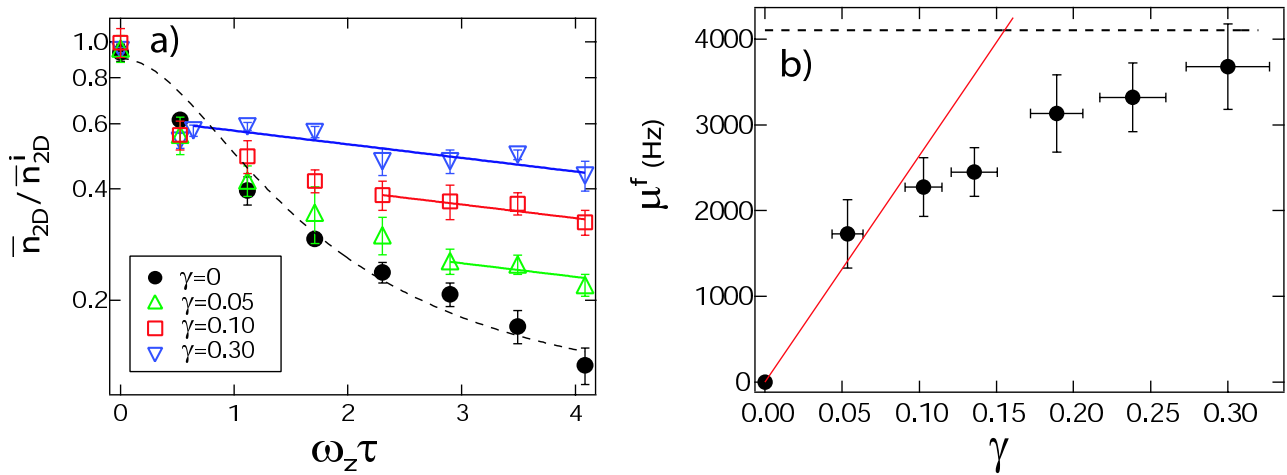}
\end{center}
\caption{\textit{\textbf{a)} Ratio $\nCORE(\tau)/\nCORE^{\  i}$ of
the average density to the initial density at the centre of the
condensate after an expansion time $\omega_z \tau$ in the 1D
magnetic guide for different amplitudes of the random potential
$\gamma$=0, 0.05, 0.10 and 0.30. The dashed line shows the
predicted time evolution according to scaling theory for
$\gamma=0$. After the onset of disorder-induced trapping, atom
losses lead to a purely exponential decay indicated by the solid
line fits. The onset of a purely exponential decay (marked by the
start of each solid line) indicates the final density $\nCORE^f$
at which disorder-induced trapping occurs. \textbf{b)} Final
effective potential $\mu^{\rm{f}}=\mu_{\rm{TF}}^{1/3}\left (
\frac{3 g \nCORE^{\ f}}{4 R_{\rm{TF}}} \right ) ^{2/3}$ at the
centre of the trapped condensate as a function of the amplitude of
the random potential $\gamma$. The (red) solid line corresponds to
the expected slope from \Eref{gamma_dependence}. The black dashed
line corresponds to the saturation value $0.92 \mu$
[\Eref{saturated_mu}].}\label{figure9}}
\end{figure}

When disorder-induced trapping occurs at the centre, the average
density in the core has dropped to a final value $\nCORE^{\  f}
\sim \gamma \mu_{\rm{TF}} / g$ [see
\Eref{gamma_dependence}-\eref{mu_nCORE}] and is then expected to
remain stationary. However, because of atom losses due to
processes such as evaporation, collisions, etc, the density in the
core will continue to fall. In \Fref{figure9}a we plot the time
evolution of the ratio of the average density $\nCORE(\tau)$ over
the initial one $\nCORE^{\ i}$ for different amplitudes $\gamma$
of the random potential. Without the random potential (black dots
$\bullet$), the time evolution of the average density is in very
good agreement with the predicted expansion in the magnetic
waveguide according to scaling theory (dashed black line). In the
presence of the random potential, the rate of decrease of the
average density $\nCORE(\tau)$ is much reduced (see
\Fref{figure9}a). In particular, once the core of the condensate
stops expanding, the evolution of $\nCORE(\tau)$ changes to an
exponential decay (the solid lines in \Fref{figure9}a). This
exponential decay is due to atom losses and indicates that the
density is no longer decreasing due to expansion. Hence the onset
of disorder-induced trapping is marked by a change of slope in the
time evolution of the density $\nCORE(\tau)$, and is indicated by
the start of each solid line in \Fref{figure9}a). For a speckle
amplitude $\gamma=0.30$, trapping occurs at $\omega_z \tau=0.5$.
The subsequent decrease in the density $\nCORE(\tau)$ ($\omega_z
\tau > 0.5$) is fitted with an exponential $\exp[-\Gamma \tau]$,
giving the time constant $1/\Gamma \simeq$ 280 ms for atom losses.
We then use an exponential with the same time constant $1/\Gamma$
to fit the curves corresponding to $\gamma=0.05$ and
$\gamma=0.10$. The onset of this exponential decay gives us a
measurement of the final average density $\nCORE^f$ for which
disorder-induced trapping occurs. In \Fref{figure9}, the error
bars on $\nCORE^f$ represent the difference in the density between
the last point considered as part of the expansion and the first
point marking the onset of disorder-induced trapping.

From the analysis of our experimental data in \Fref{figure9}a we
extract the final effective chemical potential
$\mu^{\rm{f}}=\mu_{\rm{TF}}^{1/3}\left ( \frac{3 g \nCORE^{\ f}}{4
R_{\rm{TF}}} \right ) ^{2/3}$ and plot $\mu^{\rm{f}}$ versus the
speckle amplitude $\gamma$ in \Fref{figure9}b. Comparing the final
effective chemical potential $\mu^{\rm{f}}$ once the core of the
BEC is trapped with the amplitude $\gamma$ of the random potential
allows a test of \Eref{gamma_dependence}-\eref{saturated_mu}.
Indeed, according to \Eref{gamma_dependence}, the slope of the
function $\mu^{\rm{f}}(\gamma)$ for small values of $\gamma$
reflects the first and second-order statistical properties of the
disorder, in particular the exponential one-point distribution of
intensity of the speckle and the correlation length $\Delta z$.
For the parameters of our speckle potential we expect to obtain
for the asymptotic value of the effective chemical potential
$\mu^{\rm{f}} \simeq 26.4(2) ~ 10^3 \times \gamma$ [see
\Eref{gamma_dependence}]. The evolution of $\mu^{\rm{f}}$ at small
values of $\gamma$ is in good agreement with this predicted value
for the slope (red solid line on \Fref{figure9}b). For larger
amplitudes $\gamma$ of the random potential, $\mu^{\rm{f}}$
saturates at a nearly constant value in agreement with
\Eref{saturated_mu} (black dashed line in \Fref{figure9}b is the
expected saturation value). We note that this clearly
distinguishes between our case and the case of a lattice. In the
mean-field regime with the healing length smaller than the lattice
spacing, the expansion of the condensate in a lattice is never
suppressed as no large peak can provide a sharp stopping
\cite{Laurent_paper}. However, the decrease of the average density
of the BEC is stopped when the effective chemical potential $\mu$
equals the depth of the lattice $V$ (fragmentation). The
dependence of the final effective chemical potential
$\mu^{\rm{f}}$ with $\gamma=V/\mu_{\mathrm{TF}}$ in the case of a
lattice is then $\mu^{\rm{f}} = \mu_{\rm{TF}} \times \gamma \simeq
4.6 ~10^3 \times \gamma$, independently of the lattice spacing.
\newline

The self-averaging property of our system appears in this
measurement once again. Experimentally we measure $\nCORE^{\ f}$
which depends on $\sigma_V$ [see
\Eref{gamma_dependence}-\eref{mu_nCORE}]. The average speckle
amplitude $\sigma_V$ is related to the second order moment $m_2$
of the random potential. In \Sref{subsection_self_averaging} we
showed that the standard deviation of moment $m_2$ is expected to
vary as \Eref{Eq:asympto}. For our optical apparatus [$\Delta z =
0.95(7)$ $\mu$m], the deviation $\sigma_{m_2}(L_{\rm{TF}})$ is
less than 8\% from one realization of the speckle potential to
another \footnote{For the quasi-1D setup [$\Delta z = 5.2(2)$
$\mu$m] of \cite{Clement_PRL2005}, the condensate half-length
$L_{\rm{TF}}$ covers about 40 peaks of the random potential. We
have $\pi L_{\rm{TF}}/\Delta z \simeq 90$ and \Eref{Eq:asympto}
predicts $\sigma_{m_2}(L_{\rm{TF}})$ $\simeq$ 15\%.}. This
variation is less than the experimental shot-to-shot variations of
$\simeq 15 \%$ on $\nCORE^{\  f}$ obtained with one realization of
the speckle potential. The arguments of
\Sref{subsection_self_averaging} are therefore in agreement with
our observations and we conclude that our system can be considered
as self-averaging given our experimental resolution.

\section{Conclusion}

In conclusion, we have observed the suppression of transport of a
coherent matter-wave of interacting particles in a 1D optical
random potential. Using laser speckle patterns to create the
random potential is particularly interesting as all statistical
properties can be controlled accurately. We have developed a
technique to calibrate the average amplitude $\sigma_V$ of our
random potential using the atoms as a sensor. The spatial
correlation length $\Delta z$ is also carefully calibrated. In
addition, we have observed and justified that our experimental
conditions are such that our system is self-averaging, which is of
primary importance when studying properties of disorder. We have
extended the disorder-induced trapping scenario for an expanding
1D BEC proposed in \cite{Laurent_paper,Clement_PRL2005} to the
experimental 3D condensate. Our experimental results are in
excellent agreement with the prediction of this scenario where
interactions play a crucial role and this allows us to
experimentally study in detail the interplay between the
interactions and the random potential.

The theoretical scenario predicts a central role of the
interactions in the localization of a coherent matter-wave whose
density adapts to the fluctuations of the random potential ($\xi <
\Delta z$, where $\xi=0.11~\mu$m is the healing length). Contrary
to the case of non-interacting matter-waves where Anderson
localization is expected, the interplay between the interactions
and the disorder induces the trapping of the BEC when the chemical
potential has dropped to a value smaller than the amplitude of
typically two barriers. The particular statistical distribution of
the random potential modulations is reflected in the condition
necessary for trapping of the coherent matter-wave.

An interesting extension of this work would be to study the
transport properties of the BEC for smaller interactions, which
can be controlled through Feshbach resonances for example. In the
Thomas-Fermi regime but for $\xi > \Delta z$, a screening of the
random potential is expected \cite{Laurent_paper}. For even
smaller interaction (arbitrary small), Anderson localization may
occur.

\ack We thank D.~M.~Gangardt, G.~V.~Shlyapnikov, P.~Chavel and
J.~Taboury for useful discussions as well as F.~Moron for
technical help. We acknowledge support from the Marie Curie
Fellowship Programme (J.R.), the Fundaci\'{o}n Mazda para el Arte
y la Ciencia (A.V), the D\'{e}l\'{e}\-gation G\'{e}n\'{e}rale de
l'Armement, the Minist\`ere de la Recherche (ACI Nanoscience 201
and ANR NTOR-4-42586), the European Union (the FINAQS consortium
and grants IST-2001-38863 and MRTN-CT-2003-505032) and the ESF
(QUDEDIS programme). The atom optics group of the Laboratoire
Charles Fabry de l'Institut d'Optique is a member of Institut
Francilien de Recherche sur les Atomes Froids (IFRAF,
www.ifraf.org).
\newline

%%%%%%%%%%%%%%%%%%%%%%%%%%%%%%%%%%%%%

%%%%%%%%%%%%%%%%%%%%%%%%%%%%%%%%%%%%%%%%%%%%%%%%%%%%%%%%%%%%%%%%%%%%%%%%%%%%%%

\appendix

\section{Calculation of $\sigma_{m_2}(d)$}\label{Appendix_calculation_sigma}

The ith-order moment of a single realization of the normalized
speckle field $v(z)$ calculated over a finite length $d$ is
defined as
\begin{equation}
m_i(d) ~ = ~ \frac{1}{d} \int_{-d/2}^{d/2} dz \ v^{\rm{i}}(z).
\end{equation}
Since $v(z)$ is random $m_i(d)$ is random as well. Its statistical
standard deviation $\sigma_{m_i}(d)$ for a given length $d$ is
thus:
\begin{eqnarray}\label{Eq:define_sigma}
\sigma_{m_i}^2(d) & ~ = ~ <m_i(d)^2>  - <m_i(d)>^2,
\end{eqnarray}
where $<.>$ stands for an ensemble average over the disorder.
Since the average over the disorder $<.>$ and the integration over
a finite distance $d$ commute, we can write:
\begin{eqnarray}
<m_i(d)^2> & ~ = ~  \frac{1}{d^2} \ \int_{-d/2}^{d/2} dz \
\int_{-d/2}^{d/2} dz' \ < v^i(z) v^i(z')> \\
<m_i(d)> & ~ = ~ \frac{1}{d} \int_{-d/2}^{d/2} dz \ <v^i(z)>.
\end{eqnarray}
The calculation of $\sigma_{m_2}^2(d)$ thus requires the knowledge
of the second-order correlation function $< v^2(z) v^2(z')>$ of
the intensity field. In the following we address this point
studying the statistics of the speckle pattern.
\newline

Let $A(z)$ denote the normalized amplitude of the electric field
of the light diffused by the scattering plate, {\it i.e.}
$v(z)=A^*(z)A(z)$, and $C_A(z_1-z_2)=<A^*(z_1)A(z_2)>$ the first
order correlation function for this amplitude $A(z)$. Assuming
$A(z_1)$, $A(z_2)$,...,$A(z_{2k})$ are complex Gaussian random
variables, so that one can use the so-called Vick's theorem for
those variables,
\begin{eqnarray}
\fl
<A^*(z_1)A^*(z_2)...A^*(z_{k})A(z_{k+1})A(z_{k+2})...A(z_{2k})> =
\\ \sum_{\Pi} <A^*(z_1)A(z_p)> <A^*(z_2)A(z_q)>...<A^*(z_k)A(z_r)>
,
\end{eqnarray}
where the symbol $\sum_{\Pi}$ represents a summation over the $k!$
possible permutations $(p,..., r$) of $(1, 2,..., k)$. For the
first-order and second-order correlation functions on the
intensity field $v$ we obtain:
\begin{eqnarray}
<v(z)v(z')>=1 \ + \ |C_A(z-z')|^2 \\
<v^2(z)v^2(z')>=4 \left ( 1 \ + \ 4|C_A(z-z')|^2 \ + \
|C_A(z-z')|^4 \right ). \label{Vick_law}
\end{eqnarray}

In order to obtain a simple analytic expression for $m_2(d)$ we
approximate the auto-correlation function of the normalized
speckle electric field amplitude to a Gaussian:
\begin{equation}
|<A^*(z)A(z')>|^2 = |C_A(z-z')|^2 = \exp \left [ - \left (
\frac{\pi (z-z')}{\sqrt{3} \Delta z} \right )^2 \right ].
\end{equation}

This Gaussian function has the Taylor expansion at
$(z_1-z_2)\rightarrow 0$ as the true auto-correlation function of
the speckle pattern $\sin(\pi z)/ \pi z$ for a rectangle aperture
(see \Sref{Section_speckle_field}). The calculation can also be
done using the true correlation function, but leads to a more
complex formula.
\newline

The calculation of the deviation $\sigma_{m_1}(d)$ with the
Gaussian auto-correlation function leads to the following
equation:
\begin{eqnarray}
\sigma^2_{m_1}(d) = \frac{\sqrt{3 \pi}}{u} \ \rm{Erf} \left (
\frac{u}{\sqrt{3}} \right )  +  \frac{3}{u^2} \left ( \rme^{-
u^2/3}-1 \right ). \label{Eq:calcul_m1_carre}
\end{eqnarray}
Here, $\rm{Erf}(x)=\frac{2}{\sqrt{\pi}} \int_0^{x} dt \exp(-t^2)$
is the Error function and the dimensionless variable $u=\frac{\pi
d}{\Delta z}$ is related to the typical number of speckle grains
in a given length $d$ of the system, $d/\Delta z$. Let us now turn
to the calculation of $\sigma_{m_2}^2(d)$ itself. Using
\Eref{Vick_law}, we can relate $\sigma_{m_2}^2(d)$ to
$\sigma_{m_1}^2(d)$:
\begin{equation}
\sigma_{m_2}^2(d) = 16 \sigma_{m_1}^2(d) \ + \ 4
\sigma_{m_1}^2(\sqrt{2}d). \label{Eq:relation_m1_m2}
\end{equation}
Substituting \Eref{Eq:calcul_m1_carre} into
\eref{Eq:relation_m1_m2}, we obtain an analytic expression for
$\sigma_{m_2}(d)$.  We have verified numerically that this result
is a very good approximation to that obtained using the true
auto-correlation function to within a few percent, as shown in
\Fref{figure6}. In the asymptotic limit $d \gg \Delta z$, we
obtain:
\begin{equation}\label{Eq:asympto_sigma_m2}
\sigma_{m_2}(d) \simeq \left ( \frac{2 \sqrt{3 \pi}}{u}
(8+\sqrt{2}) \right )^{1/2} \simeq 7.60 \ \frac{1}{\sqrt{u}}.
\end{equation}
In the opposite limit $d \rightarrow 0$, we find
\begin{equation}
\sigma_{m_2}(d=0)=\sqrt{20} \label{sigma_m2_zero}.
\end{equation}

The asymptotic function \Eref{Eq:asympto_sigma_m2} is a very good
approximation of the analytical solution even for small numbers of
peaks $d/\Delta z$. Indeed the difference between the asymptotic
and analytic solution is less than 1\% for systems larger than
$d/\Delta z \simeq 6$.

\section{Condensate expansion in quasi-1D and true 1D random potentials}
\label{Appendix_1D_quasi1D}

\begin{figure}[ht!]
\begin{center}
\includegraphics[width=14cm]{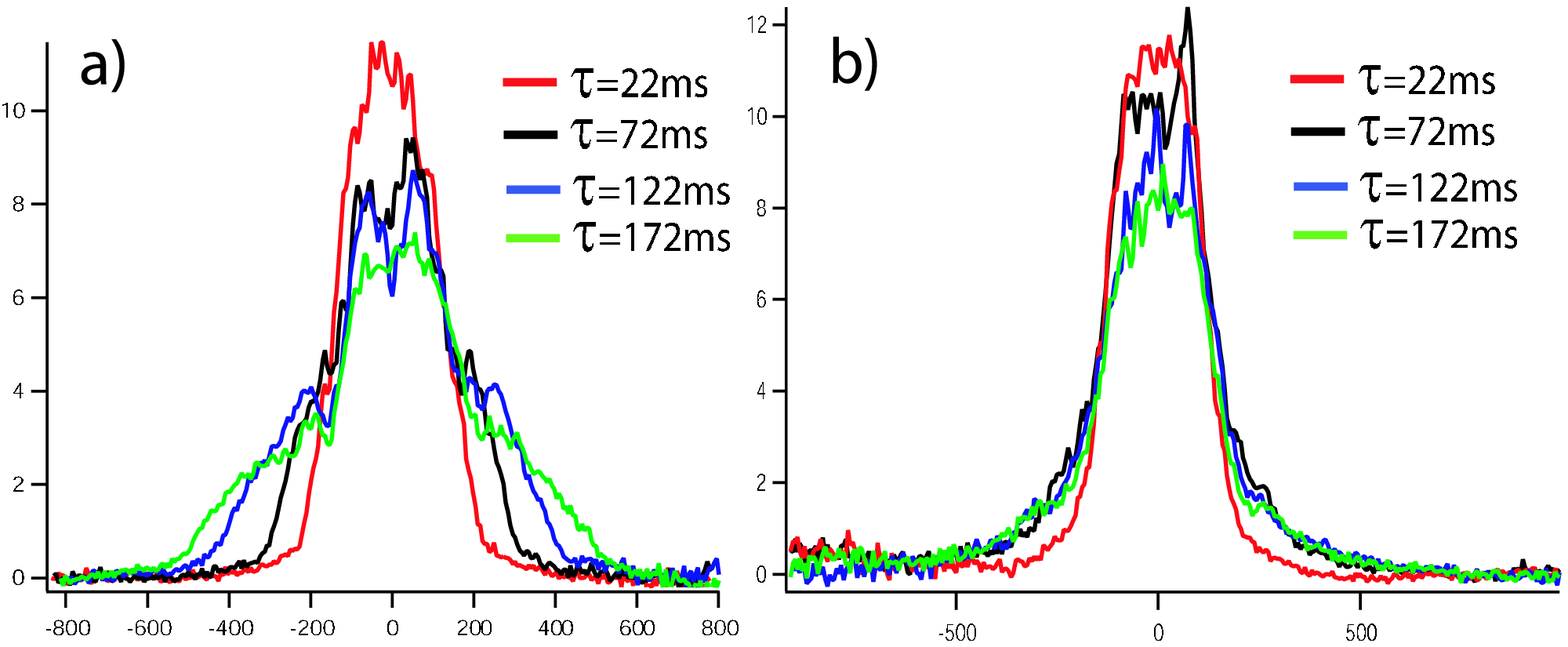}
\end{center}
\caption{\textit{Longitudinal density profiles, each averaged over
10 realizations of the speckle pattern, observed after expansion
times $\tau=$ 22, 72, 122 and 172\,ms in a speckle potential of
amplitude $\gamma = \sigma_V / \mu_{\rm{TF}} = 0.3$ for:
\textbf{a)} a quasi-1D potential ($\Delta y / R_{\rm{TF}}\simeq
3.5$) and \textbf{b)} a true 1D potential ($\Delta y /
R_{\rm{TF}}\simeq 36$). In a), parabolic ``wings''' are visible
around the central, trapped core. These continue to expand as a
function of time.}\label{figure10}}
\end{figure}

In this article, we have presented a comparison of the expansion
of a condensate in a 1D random potential with a theoretical
scenario, obtaining good quantitative agreement.  However, we have
found that the 1D requirement for the potential is quite
stringent. As described in section
\Sref{subsection_speckle_grain_size}, we are able to vary both the
size and anisotropy of the speckle grains. By performing the same
experiments with a {\it quasi-1D} potential ($\Delta y /
R_{\rm{TF}}\simeq 3.5$, as used in ref.~\cite{Clement_PRL2005}),
we observe a behaviour on the longitudinal density profiles which
is qualitatively different from that obtained using the {\it true
1D} potential ($\Delta y / R_{\rm{TF}}\simeq 36$) presented in
this article.

The main difference is the appearance of ``wings'' in the
longitudinal density profiles, which can be clearly seen in figure
\Fref{figure10}(a). The shape of these wings is approximately an
inverted parabola. Whereas the central core of the profile is
trapped by the disorder, the wings continue to expand at a rate of
3.6(1)\,mm\,s$^{-1}$.  This is significantly slower than
condensate expansion in the absence of disorder,
6.1(1)\,mm\,s$^{-1}$. To exclude the possibility of these wings
being composed of thermal atoms, produced by heating of the
condensate in the speckle potential, we repeated the experiment
with an atom cloud at 600\,nK with a condensate fraction of only
15\%.  In this case, the large thermal fraction leads to wings
with a gaussian profile, which expand with a velocity of
11(2)\,mm\,s$^{-1}$.  From this we conclude that the additional
wings appearing in the BEC during expansion in a quasi-1D random
potential must be related to possibility of condensate atoms
passing around some of the speckle peaks and continuing to expand.

The measurement of the rms size $L$ in quasi-1D potentials reveals
the phenomenon of suppression of transport as $L$ saturates
\cite{Clement_PRL2005}. Yet, since the additional wings contribute
to the rms size $L$ of the condensate, $L$ may continue to
increase for some time after the onset of disorder-induced
trapping of the central core and thus may not give a direct access
to the time-scale of the trapping scenario. Therefore, in a
quasi-1D potential, it is important to study the time evolution of
the density profiles in order to correctly obtain the predicted
timescales for this trapping phenomenon.

\end{document}